\documentclass[aps,showpacs,prl,twocolumn]{revtex4}

\usepackage{amssymb}
\usepackage{graphpap}
\usepackage[dvips]{graphicx}
\usepackage[dvips]{graphics}
\usepackage{color}

\begin{document}

\title{Spin States in Graphene Quantum Dots}
\author{J. G\"uttinger}
\email{guettinj@phys.ethz.ch}
\author{T. Frey}
\affiliation{Solid State Physics Laboratory, ETH Zurich, 8093 Zurich, Switzerland}
\author{C. Stampfer}
\altaffiliation{Present address: JARA-FIT and II. Institute of Physics, RWTH Aachen, 52074 Aachen, Germany}
\affiliation{Solid State Physics Laboratory, ETH Zurich, 8093 Zurich, Switzerland}
\author{T. Ihn}
\affiliation{Solid State Physics Laboratory, ETH Zurich, 8093 Zurich, Switzerland}
\author{K. Ensslin}
\affiliation{Solid State Physics Laboratory, ETH Zurich, 8093 Zurich, Switzerland}

\date{ \today}
 
\begin{abstract}
We investigate ground and excited state transport through small ($d \approx 70$~nm) graphene quantum dots. The successive spin filling of orbital states is detected by measuring the ground state energy as a function of a magnetic field. For a magnetic field in-plane of the quantum dot the Zemann splitting of spin states is measured. The results are compatible with a g-factor of 2 and we detect a spin-filling sequence for a series of states which is reasonable given the strength of exchange interaction effects expected for graphene. 
\end{abstract}

\maketitle

Spin qubits in quantum dots~\cite{los98} are interesting candidates for the implementation of future quantum information processing. Single spin preparation, manipulation and read-out has so far been demonstrated predominately in GaAs-based systems~\cite{elz04,pet05}.
Spin coherence times in such systems are limited by hyperfine coupling and spin-orbit interactions. In graphene-based nanostructures both limitations are expected to be significantly reduced in strength~\cite{tra07,kan05,min06,hue06}. The electostatic tunability of graphene quantum dots as well as the observation of excited states have been demonstrated recently~\cite{sta08a, pon08, liu09, sch09, mos09}. The g-factor in graphene has been measured via conductance fluctuations~\cite{lun09}, but the general spin properties of confined electrons in graphene have remained elusive. 

\begin{figure}\centering\centering
\includegraphics[draft=false,keepaspectratio=true,clip,width=0.9\linewidth]%
                   {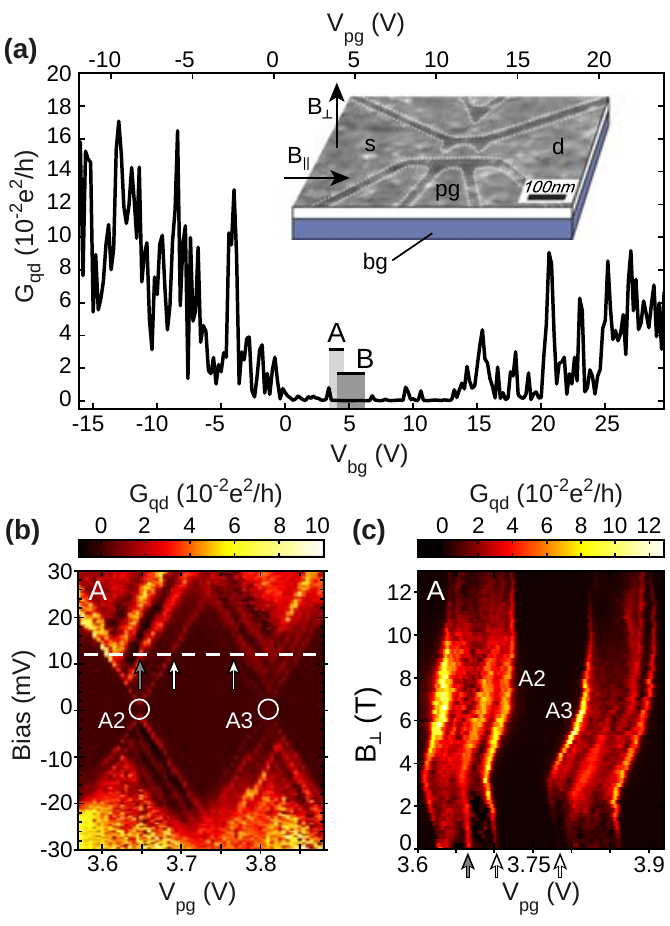}                   
\caption[FIG1]{(Color online) (a) Quantum dot source-drain conductance for varying back gate voltage ($V_{\mathrm{b}} = 100~\mu$V). The inset shows a scanning force microscope picture of the measured graphene quantum dot device. (b) Coulomb diamond measurement recorded on the hole side of the electron-hole crossover. The gate dependence at $V_\mathrm{b} = 11$~mV (white dashed line) for increasing perpendicular magnetic field is plotted in (c). Both edges of the diamond in (b) show similar magnetic field dependences typical for spin pairs.
}
\label{transport}
\end{figure}

Here we present transport measurements on a graphene quantum dot as a function of magnetic field and analyze the evolution of Coulomb peaks and excited states in both perpendicular and parallel magnetic fields. 

A scanning force microsope (SFM) picture of the sample is shown as an inset in Fig.~1(a). The graphene quantum dot in the center is connected via tunneling constrictions to source (s) and drain (d) contacts. It is tunable by the back gate (bg) and several in-plane gate electrodes, including the plunger gate (pg). 
The fabrication of the sample is described in detail in Ref.~\onlinecite{gue09b}. All measurements were performed in a dilution refrigerator with a base temperature of $90$~mK.

The source-drain conductance of the device measured over a large back gate voltage range is shown in Fig.~1(a). The region of suppressed conductance between 0 and 13~V ("transport gap"~\cite{sta09}) separates the hole-transport regime at low $V_{\mathrm{bg}}$ from the electron transport regime at high $V_{\mathrm{bg}}$. Coulomb-blockaded transport through the quantum dot is observed in this region at closer inspection. Regions A and B indicate the gate voltage ranges of the two measurement regimes investigated in this paper. Previous measurements in region A~\cite{gue09} gave evidence for an electron-hole crossover in the Coulomb-blockaded quantum dot around $V_{\mathrm{bg}} = 5.5$~V.

The same conductance resonances scanned with the back gate in region A (B) can also be investigated by keeping $V_{\mathrm{bg}}$ fixed at -0.9~V (-1.8~V) and sweeping the plunger gate voltage [see top-axis in Fig.~1(a) for the calculated plunger gate voltage at $V_{\mathrm{bg}} = -0.9$~V (region A) corresponding to the measured back gate voltage. This plunger gate scale has been used throughout the paper as a gate voltage scale to ease comparison].

Two Coulomb-blockade resonances from regime A are investigated by measuring the differential conductance as a function of plunger gate voltage and source-drain bias at $B = 0$~T in Fig.~1(b). At $V_{\rm{b}}=0$~mV the differential conductance is suppressed except for the Coulomb resonance conditions (circles) where two conductance peaks A2 and A3 are observed. By increasing the bias window a diamond shaped region of blocked conductance is measured, revealing an electron addition energy of 20~meV. The change in differential conductance parallel to the edges of the diamond is attributed to excited states providing additional transport channels. Their energy splitting from the ground state is obtained directly from the measurement (e.g. 6~meV for the state marked with the gray arrow).

A cut through the diamond at fixed bias voltage $V_{\mathrm{b}} = 11~$mV [dashed white line in Fig.~1(b)] for increasing magnetic field perpendicular to the graphene plane is shown in Fig.~1(c). After a kink around $B = 3.5~$T the edges of the diamonds move towards higher gate voltages with increasing field. This kink has been interpreted in Refs.~\onlinecite{gue09,lib10} as an indication of filling factor two for holes in the quantum dot. The same kink is observed for the excited state resonance at $V_{\mathrm{pg}} = 3.66$~V. The comparable evolution of the two diamond edges (white arrows) could be a result of spin pairing, i.e., two subsequently filled electrons occupying the same orbital state with opposite spin orientation.

Such pairs are interesting candidates for studying spin properties. In general, the extraction of a small Zeeman splitting from the magnetic field dispersion of individual conductance resonances is difficult, because any small orientational misalignment of the sample causes a significant perpendicular magnetic field component. Typically, the dispersion of states in the perpendicular field dominated by orbital effects is much stronger than the faint Zeeman splitting. Hence, small misalignment severely hampers the extraction of precise g-factor values from the dispersion of individual peaks. The problem can be circumvented by analyzing the peak-to-peak spacing of spin-pairs, because orbital contributions can be significantly reduced by subtracting the positions of individual peaks sharing the same orbital shift~\cite{Tarucha96, lin02}.

Potential spin pairs are also identified in the gate regime B. In Fig.~2(a) the evolution of Coulomb peaks with increasing perpendicular field is displayed. The lowest two peaks (B1, B2) and the following two (B3, B4) are identified as potential spin pairs due to their similar peak evolution. In Fig.~2(b) a measurement of the same peaks is shown after careful in-situ rotation of the sample into an orientation parallel to the magnetic field. 

\begin{figure}\centering
\includegraphics[draft=false,keepaspectratio=true,clip,width=1.0\linewidth]%
                   {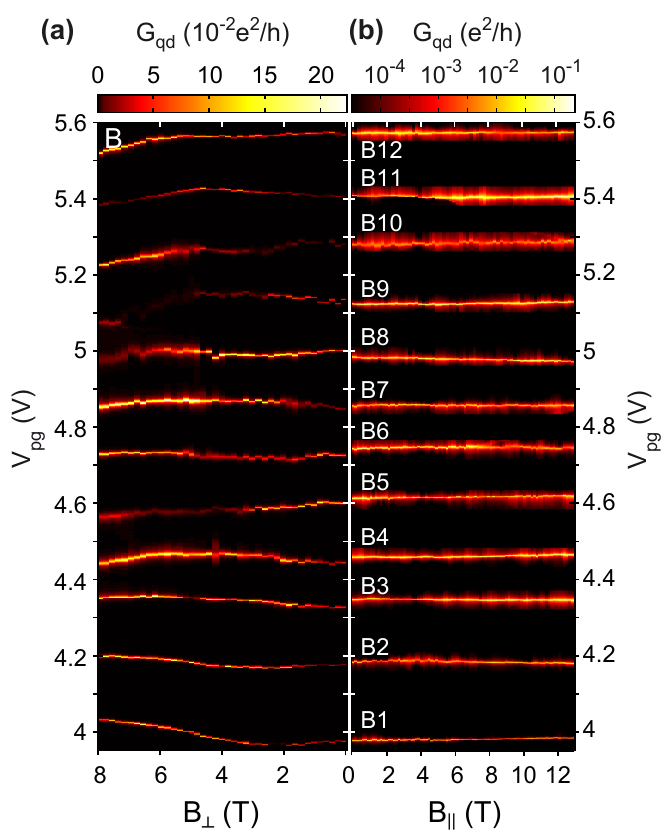}                   
\caption[FIG2]{(Color online) (a) Coulomb peaks as function of perpendicular magnetic field recorded at $V_{b} = 100~\mu$V. The conductance is plotted in linear scale to enhance visibility of amplitude changes. (b) Same Coulomb peaks measured in parallel (in-plane) magnetic field after tilting the sample by $90^\circ$. Here the conductance is plotted in logarithmic scale to increase the visibility of the peaks.} 
\label{experiment}
\end{figure}

\begin{figure*}\centering
\includegraphics[draft=false,keepaspectratio=true,clip,%
                   width=1.0\linewidth]%
                   {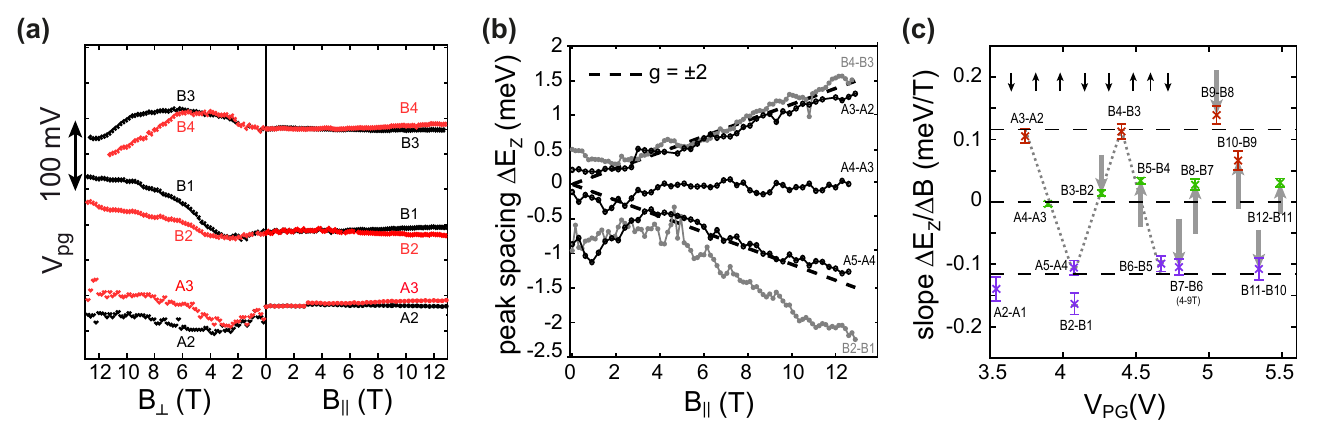}                   
\caption[FIG3]{(Color online) (a) Comparing the evolution of three peak pairs in perpendicular (left) and parallel (right) magnetic field. (b) Coulomb peak spacing as function of $B_{\parallel}$ for the three pairs shown in (a) and the two subsequent peak spacings (A4-A3 and A5-A4) in regime A. The dashed lines show the Zeeman splitting $\Delta E^\mathrm{Z} = \pm|g|\mu_\mathrm{B} B$ for a g-factor $|g| = 2$. All peaks are converted to energy using the lever arm $\alpha_{\mathrm{pg}} = 0.13\pm 0.01$ as extracted from the corresponding Coulomb diamond measurements. (c) Slopes obtained by a linear fit for $B_{\parallel} = 4.2-13$~T of peak-to-peak spacings shown in (b) as a function of $V_{\mathrm{pg}}$. The slopes of measurement B are compensated for a parallel magnetic field misalignment of $\beta_\mathrm{B}=3^\circ$~ (see text), where the shift induced by the compensation is depicted by the gray arrows. The peak difference B7-B6 has a strong kink at $B_{\parallel} \approx 9$~T and is fitted only up to 9~T. The spin-filling-sequence on top is extracted from the points along the dotted gray line.} 
\label{analysis}
\end{figure*}

To analyze the movement of the peaks shown in Fig.~2, the peak positions are extracted by fitting. The results for the two pairs (B1-B4) in Fig.~2 and the pair (A2-A3) in Fig.~1(b,c) are shown in Fig.~3(a). Peak positions are plotted with suitable offsets in $V_{\mathrm{pg}}$ such that pairs coincide at $B=0$~T. For low perpendicular magnetic field ($B_\perp < 3$~T for A2/A3, B1/B2; $B_\perp < 6$~T for B3/B4) the orbital states of each pair have approximately the same magnetic field dependence. Hence spurious orbital contributions to the peak spacing resulting from slight misalignment are limited. The data in Fig.~3(a) for parallel orientation of the magnetic field shows roughly linear splitting of pairs with increasing $B_{\parallel}$. We interpret this splitting as the Zeeman splitting of two spin states with identical orbital wave functions.

The magnitudes of the Zeeman splitting of the three pairs discussed above and for two additional peak-spacings from region A are plotted in Fig.~3(b). The data is obtained in two steps. First, the peak-to-peak spacing from Fig.~2(b) and corresponding measurement in regime A is extracted. Second, the peak spacing is linearly fit from 4.2~T to 13~T and the offset obtained from the fit is subtracted. The Zeeman spin splitting $\Delta E^\mathrm{Z}$ is given by \cite{fol01,lin02}
\[
\Delta E_{N+1}^{\mathrm{Z}} = \left(S_{N+1}-2S_{N}+S_{N-1}\right)g\mu_{\mathrm{B}} B_{\parallel} + \textrm{const.}
\]
Here $S_{N}$ is the spin quantum number along the $B_{||}$-direction of the dot with $N$ electrons. Spin differences of successive ground states $\Delta S_{N+1}^{\mathrm{(1)}} = S_{N+1}-S_{N}$ are half-integer values (e.g. $\Delta S^{\mathrm{(1)}} = 1/2$ for adding a spin up electron or $\Delta S^{\mathrm{(1)}} = 3/2$ for adding a spin up electron while "flipping" another spin from down to up). Hence differences between three successive spin ground states take on integer values $\Delta S_{N+1}^{\mathrm{(2)}} = S_{N+1}-2S_{N}+S_{N-1} = 0, \pm 1,\pm 2, \ldots$. Apart from the splitting of the pair B2-B1 the splitting of all pairs in Fig.~3(b) show slopes $\Delta E^{\mathrm{Z}} = \Delta S^{\mathrm{(2)}}g\mu_B B_{||}$ with integer values $\Delta S^{\mathrm{(2)}} =  0, \pm 1$ and a g-factor value of approximately 2 [black dashed lines in Fig.~3(b)]. 

We now attempt to extract spin-related shifts from individual unpaired conductance resonances despite the inherent difficulty mentioned above. The final goal is to reconstruct a plausible spin-filling sequence of the quantum dot in regimes A and B. We apply the following strategy: first, the misalignment angle $\beta$ is estimated, such that second, the peak shifts in parallel field can be corrected for the orbital effects arising from the misalignment.
Different angles $\beta$ exist for measurements in regimes A and B, because the sample had been rotated into perpendicular orientation between the measurements B and A. The misalignment angle $\beta$ was estimated assuming $g=2$ and linear peak shifts in $B_\perp$ in the range $0\leq B_\perp\leq 2\,$T and $B_\parallel$ in the range $4.2\,\mbox{T}\leq B_\parallel\leq 13\,\mbox{T}$. The angle $\beta$ was then chosen to generate approximately $\Delta S_{N+1}^{(1)}=\pm 1/2$ for all peaks considered. The obtained values are $\beta_\mathrm{A}=0^\circ \pm 0.5^\circ$ and $\beta_\mathrm{B}=3^\circ \pm 1^\circ$. As a consequence no correction for orbital effects was necessary in regime A.

The misalignment-compensated-slopes of the peak-spacings are shown in Fig.~3(c) as a function of the plunger gate voltage $V_{\mathrm{pg}}$. The gray arrows indicate the change of the slope by the compensation of the angular misalignment. The error bar includes the error from both the lever arm estimation (10\% of the slope) and the variance from the linear fit of the peak-to-peak spacings. Three dashed black lines indicate the slopes corresponding to $\Delta E^{\mathrm{Z}}/\Delta B = \{0,\pm1\}\times2\mu_{\mathrm{B}}$ and the data points are colored according to their closest $\Delta S^{\mathrm{(2)}}$ value (blue for $\Delta S^{\mathrm{(2)}}=-1$, green for $\Delta S^{\mathrm{(2)}}=0$ and red for $\Delta S^{\mathrm{(2)}}=+1$). Note that the two regimes A and B overlap: pairs B2-B1 and A5-A4 are in fact the same Coulomb peaks as it can be seen from a conductance measurement in dependence of $V_\mathrm{pg}$ and $V_\mathrm{bg}$ (not shown). 
We can extract the spin-filling-sequence $\downarrow\uparrow\uparrow\downarrow\downarrow\uparrow\uparrow\downarrow$ from peak A2 to B6 (dotted gray line). 
We emphasize that we see a clear deviation from $\uparrow\downarrow\uparrow\downarrow$ as it has been seen in the low carrier regime of carbon nanotube quantum dots~\cite{bui02,cob02,JaHe04,mor05}.

This result can be made plausible by considering the exchange interaction between the charge carriers. The strength of the Coulomb interaction in graphene can be characterized by the ratio $r_{\mathrm{s}}$ between interaction energy and kinetic energy~\cite{hei07}
\[
r_{\mathrm{s}} = \frac{E_{\mathrm{int}}}{E_{\mathrm{F}}} = \frac{e^2/(4\pi\epsilon_0\epsilon_{\mathrm{r}} r_{\mathrm{ee}})}{\sqrt{\pi\hbar^2v_{\mathrm{F}}^2 n_{\mathrm{s}}}} = \alpha_{\mathrm{g}} \approx 0.9,
\]
independent of the charge carrier density. Here $n_{\mathrm{s}}$ is the density of states, $r_{\mathrm{ee}} = (\pi n_{\mathrm{s}})^{-1/2}$ is the mean electron separation, $v_{\mathrm{F}}$ the Fermi velocity in graphene and $\alpha_{\mathrm{g}} = \alpha \cdot v_{\mathrm{F}}/\epsilon_{\mathrm{r}}c$ the \textit{graphene fine-structure constant}. For the dielectric constant a mixture of vacuum and SiO$_2$ $\epsilon_{\mathrm{r}} = 2.5$ is used. The splitting of the eigenstate energies due to the exchange interaction can be estimated using $\xi(r_{\mathrm{s}} = 1) \approx 0.6\Delta$~\cite{bla97, pat98, lue01}. The exchange energy splitting is comparable to the single-particle-level-spacing $\Delta$ and can therefore lead to ground-state spin polarization $S > 1/2$ in agreement with our interpretation and observations in GaAs quantum dots~\cite{fol01, lin02}. An alternative reason for spin-polarization in graphene could arise from the valley degeneracy. However, our measurements show no indications of four-fold shell filling as it has been seen in carbon nanotubes~\cite{bui02,cob02,mor05}. 

\begin{figure}\centering
\includegraphics[draft=false,keepaspectratio=true,clip,%
                   width=1.0\linewidth]%
                   {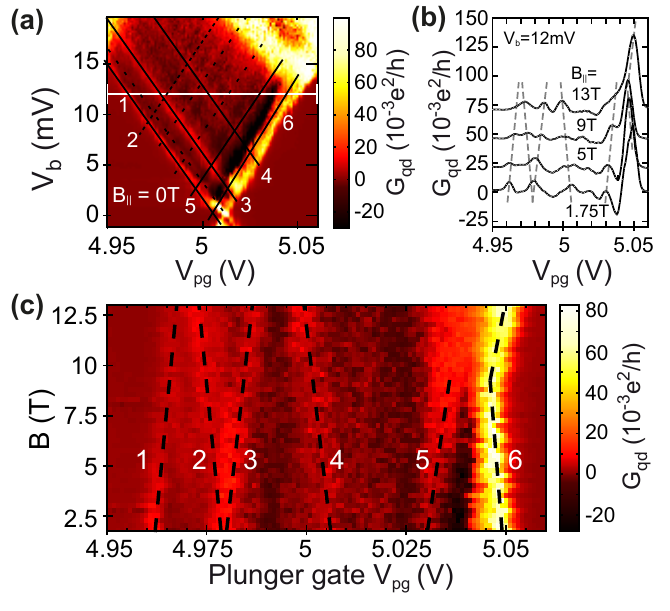}                   
\caption[FIG4]{(Color online) (a) Close up of a Coulomb diamond measurement around the electron-hole crossover at $B_{\parallel} = 0$~T. (b,c) Cross section at $V_\mathrm{b} = 12$~mV [white line in Fig.~3(a)] for increasing in-plane magnetic field. In (b) the traces are offset in conductivity and smoothed over a five point window for clarity. The dashed lines in (c) have two different slopes corresponding to a Zeeman spin splitting with $g=2$ and are plotted for comparison over the raw data. } 
\label{diapara}
\end{figure}

Next we describe an attempt to observe the Zeeman splitting of excited states in parallel magnetic field. Fig.~4(a) shows part of a Coulomb blockade diamond measurement at $B = 0$~T (this is a close-up of Fig.~2(b) in Ref.~\onlinecite{gue09}). As in Fig.~1(b) several differential conductance peaks are visible. Along the white bar at $V_\mathrm{b} = 12~$mV the evolution of the peaks is measured in Fig.~4(c). Individual cross sections are plotted in Fig.~4(b) showing the change in amplitude and the movement of the peaks. 
The six peaks labeled in Fig.~4(c) are linked to the stronger resonances in Fig.~4(a). The magnetic field dependence of the two faint lines between resonance 2 and 5 in Fig.~4(a) could not be traced in Fig.~4(c). 
The evolution of the peaks can be attributed to two different slopes. For comparison black dashed lines with slopes differing by the Zeeman energy $\Delta E^\mathrm{Z} = |g|\mu_\mathrm{B} B$ with $g=2$ are plotted on top. Despite a slight underestimation of the splitting, the lines agree reasonably well with the measurement. Hence we attribute the two slopes to Zeeman spin splitting. The absence of clear peak splittings that would be induced by lifting the spin degeneracy with B can be seen as another indication of significant charge carrier interactions. 

In summary, we have analyzed a graphene quantum dot in an in-plane magnetic field by investigating the conductance peak spacing and the evolution of excited states around the electron-hole crossover. In both regimes A and B spin-pairs were identified showing a linear and $|g|=2$ compatible Zeeman splitting with magnetic field, whereas other conductance peaks are difficult to analyze similar to investigations in GaAs quantum dots~\cite{fol01, lin02}. The observed spin sequence with spin polarization and the absence of clear degenerate ground and excited states are in agreement with the strength of Coulomb interactions expected in graphene. 

The authors wish to thank J.~Seif, P.~Studerus, C.~Barengo, M.~Csontos, Y.~Komijani, B.~K\"ung, T.~Helbling and T.~Heinzel for help and discussions. 
Support by the ETH FIRST Lab, the Swiss National Science Foundation and NCCR nanoscience are gratefully acknowledged.

\end{document}